\documentclass[%
reprint,
superscriptaddress,
bibnotes,longbibliography
amsmath,amssymb,
aps,
floatfix
]{revtex4-2}
\usepackage[dvipdfmx]{graphicx}
\usepackage{graphicx}
\usepackage{color}
\usepackage{natbib}
\usepackage{amsmath,amssymb,amsthm,mathrsfs,amsfonts,dsfont}
\usepackage{subfigure, epsfig}
\usepackage{braket}
\usepackage{bm}
\usepackage{bbm}
\usepackage{enumerate}
\usepackage{comment}
\usepackage{appendix} 
\usepackage[colorlinks,linkcolor=blue,citecolor=blue]{hyperref}
\newcommand{\tr}{\mathrm{tr}}

\usepackage{makecell} 

\newtheorem{theorem}{Theorem}

\theoremstyle{definition}

\theoremstyle{remark}

\theoremstyle{example}

\newcommand{\polylog}{\mathrm{polylog}}

\usepackage[utf8]{inputenc}
\usepackage{array}       
\usepackage{cellspace}   
\usepackage{graphicx}    

\newcolumntype{C}[1]{>{\centering\arraybackslash}m{#1}}
\addparagraphcolumntypes{C}   

\begin{document}
\title{Testing the equivalence to thermal states via extractable work under LOCC}

\author{Toshihiro Yada}
\email{yada@noneq.t.u-tokyo.ac.jp}
\affiliation{\mbox{Department of Applied Physics, The University of Tokyo, 7-3-1 Hongo, Bunkyo-ku, Tokyo 113-8656, Japan}}

\author{Nobuyuki Yoshioka}
\affiliation{\mbox{International Center for Elementary Particle Physics, The University of Tokyo, 7-3-1 Hongo, Bunkyo-ku, Tokyo 113-0033, Japan}}

\author{Takahiro Sagawa}
\affiliation{\mbox{Department of Applied Physics, The University of Tokyo, 7-3-1 Hongo, Bunkyo-ku, Tokyo 113-8656, Japan}}



\begin{abstract}
Understanding the thermal behavior of quantum many-body pure states is one of the most fundamental issues in quantum thermodynamics.
It is widely known that typical pure states yield vanishing work, just as thermal states do, when one restricts to local operations that cannot access correlations among subsystems. 
However, it remains unclear whether this equivalence to thermal states persists under LOCC (local operations and classical communication), where classically accessible correlations can be exploited for work extraction.
In this work, we establish criteria for determining whether many-body pure states remain equivalent to thermal states even under LOCC, and show that this thermal equivalence is governed by their multipartite quantum correlation structure.
We show that states with asymptotically maximal multipartite entanglement, such as Haar-random states, cannot yield extensive work under LOCC, whereas some states with limited multipartite entanglement, such as constant-degree graph states, allow extensive work extraction despite being locally indistinguishable from thermal states.
Thus, our work provides a refined operational notion of thermal equivalence beyond the traditional local regime, which is becoming increasingly important due to the recent expansion of experimentally accessible operations.
\end{abstract}
\maketitle

\noindent
\textbf{Introduction:}
How thermal behavior emerges in isolated quantum many-body systems is one of the central problems in quantum thermodynamics~\cite{d2016quantum,gogolin2016equilibration,deutsch2018eigenstate,mori2018thermalization,binder2018thermodynamics,ueda2020quantum}. 
This problem dates back almost a century to von Neumann's seminal work~\cite{von2010proof} and has since been intensively investigated~\cite{trotzky2012probing,langen2013local,kaufman2016quantum,shaw2025experimental,google2025observation,rigol2008thermalization,beugeling2014finite,kim2014testing}.
A key to understanding this phenomenon is the restriction to the local regime: 
typical pure states are indistinguishable from thermal states when one focuses on reduced density operators of local subsystems~\cite{popescu2006entanglement,goldstein2006canonical,reimann2007typicality,tasaki2016typicality}. 

This local equivalence can also be characterized operationally in terms of work extraction~\cite{oppenheim2002thermodynamical,zurek2003quantum,horodecki2005local,synak2004classical,devetak2005distillation,brodutch2010quantum,braga2014maxwell,yada2024measuring}. 
When only local operations are allowed, the extractable work from typical pure states vanishes, as it does for thermal states, because local operations cannot utilize correlations among subsystems. 
By contrast, when arbitrary global operations are allowed, extensive work can be extracted from any pure state, since one can fully exploit the multipartite correlations. 

\begin{figure}[]
    \centering
    \includegraphics[width=0.5\textwidth]{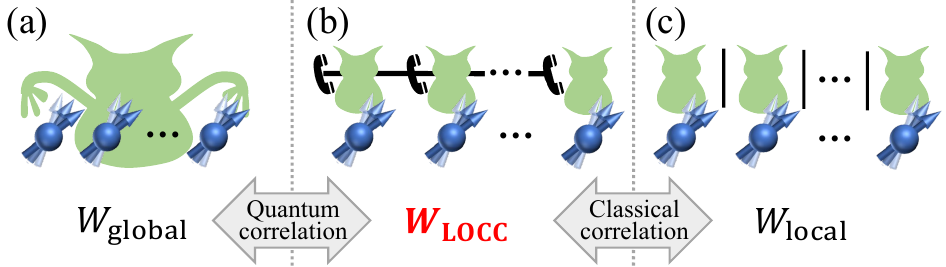}
    \caption{Schematics of work extraction under (a) arbitrary global operations, (b) LOCC, and (c) strictly local operations. The increase in extractable work from local operations to LOCC reflects the contribution of classically accessible multipartite correlations, whereas the gap between global operations and LOCC quantifies multipartite quantum correlations.}
    \label{fig:setup}
\end{figure}

This naturally raises the question of whether thermal equivalence holds between these two extremes, when the allowed operations are more powerful than strictly local ones. 
In particular, this question is highly nontrivial for LOCC (local operations and classical communication), because one can exploit classically accessible correlations for work extraction in this setup.
Importantly, LOCC provides an operational separation between classical and quantum correlations, as shown in Fig.~\ref{fig:setup}: the gap between the extractable work under LOCC and that under global operations quantifies the genuinely quantum part of multipartite correlations~\cite{oppenheim2002thermodynamical,zurek2003quantum,horodecki2005local,synak2004classical,devetak2005distillation,brodutch2010quantum,braga2014maxwell,yada2024measuring}.

\begin{table}[]
    \centering
    \begin{tabular}{cccc} 
    \hline
     State ensemble & $W_{\rm global}$ &  $W_{\rm LOCC}$ &  $W_{\rm local}$ \\ \hline\hline
    Uniform (Haar) distribution & $\Theta(N)$ & $o(N)$ & $o(1)$ \\ \hline
    Approx. $t$-design, $t=\omega (\ln N)$ & $\Theta(N)$ & $o(N)$ & $o(1)$ \\ \hline
    Random graph states & $\Theta(N)$ & $o(N)$ & $o(1)$ \\ \hline
    Constant-degree graph states  & $\Theta(N)$ & $\Theta(N)$ & $o(1)$ \\ \hline
    Subset state ensemble & $\Theta(N)$  & $\Theta(N)$ & $o(1)$ \\ \hline
    \end{tabular}
    \caption{
    Extractable work for each state ensemble in an $N$-qudit system under global operations, LOCC, and local operations. The main result in this Letter consists of the scaling of $W_{\rm LOCC}$ for each type of pure state. Here, $o(1)$ denotes a quantity that vanishes as $N\to\infty$, $o(N)$ denotes a subextensive quantity, $\Theta(N)$ denotes an extensive quantity, and $\omega(\ln N)$ denotes a quantity that grows faster than $\ln N$.
    }
    \label{tab:summary}
\end{table}

In this work, we establish criteria to determine when work extraction is dominated by multipartite quantum correlations rather than classically accessible correlations, and examine thermal equivalence under LOCC for several types of many-body pure states.
First, we show that the extractable work under LOCC becomes subextensive when a measure of multipartite entanglement called \textit{geometric entanglement}~\cite{barnum2001monotones,wei2003geometric,shimony1995degree} is asymptotically maximal.
This class of highly entangled states includes Haar-random states, random graph states~\cite{hein2006entanglement}, and states sampled from approximate state $t$-designs~\cite{mele2024introduction} with superlogarithmic $t$, as summarized in Table~\ref{tab:summary}. 

On the other hand, we demonstrate that some states with limited multipartite entanglement allow extensive work extraction under LOCC, despite their local equivalence to thermal states. 
These results are obtained by explicitly constructing tailored work extraction protocols for each state and deriving a lower bound on the extractable work.
We obtain such results for graph states with constant degree and for states sampled from the subset state ensemble~\cite{aaronson2022quantum,giurgica2023pseudorandomness,jeronimo2023pseudorandom}, which is an ensemble with limited coherence in the computational basis.
Thus, our work sharpens the notion of thermal equivalence beyond the traditional local regime, providing an operational criterion that separates pure states that are locally indistinguishable from thermal states into distinct classes based on their multipartite quantum correlation structure.

\vspace{1mm}
\noindent
\textbf{Work-extraction framework:}
We begin by introducing the work-extraction framework used throughout this Letter.
According to the second law of thermodynamics, the amount of work extractable from a system is given by 
$W = k_{\rm B} T\, \Delta S - \Delta E$, 
where $\Delta S$ is the entropy change, $\Delta E$ is the change in internal energy, and $T$ is the heat-bath temperature.
In this work, we consider the infinite-temperature regime $k_{\rm B} T \gg \Delta E$, in which the energetic contribution becomes negligible compared to the entropic term. 
This setting provides a simple situation where the extractable work is determined solely by the purity of the state, regardless of the system's Hamiltonian, and has therefore been widely employed in previous works~\cite{oppenheim2002thermodynamical,zurek2003quantum,horodecki2005local,synak2004classical,devetak2005distillation,brodutch2010quantum,braga2014maxwell,yada2024measuring}.
We note that this regime is in clear contrast with the frameworks of ergotropy~\cite{allahverdyan2004maximal} and passivity~\cite{pusz1978passive,lenard1978thermodynamical,skrzypczyk2015passivity}, which focus exclusively on internal energy changes $\Delta E$ \cite{perarnau2015extractable,huber2015thermodynamic,bruschi2015thermodynamics,andolina2019extractable,elouard2017role,manzano2018optimal,kaneko2019work,baba2023work,hokkyo2025universal}. 

In this regime, the work extractable from an $N$-qudit state $\rho$ under global operations is given by \cite{alicki2004thermodynamics,jacobs2009second,sagawa2022entropy}
\begin{equation}
\label{eq:W_global}
    W_{\rm global}(\rho) = N \ln d - S(\rho),
\end{equation}
where $d$ is the local dimension and we set $k_{\rm B} T=1$ for simplicity.
This expression represents the entropy change when the initial state $\rho$ is transformed into the maximally mixed state $(\mathbbm{1}/d)^{\otimes N}$.
On the other hand, when operations are restricted to strictly local ones, the extractable work becomes \cite{synak2004classical,devetak2005distillation,modi2012classical}
\begin{equation}
\label{eq:W_local}
    W_{\rm local}(\rho) = N \ln d - \sum_{n=1}^N S(\rho_n),
\end{equation}
where $\rho_n = \tr_{\overline{n}}[\rho]$ is the reduced density operator of the $n$-th subsystem.
This is interpreted as the sum of the extractable work from each local subsystem.

\begin{figure}[]
    \centering
    \includegraphics[width=0.5\textwidth]{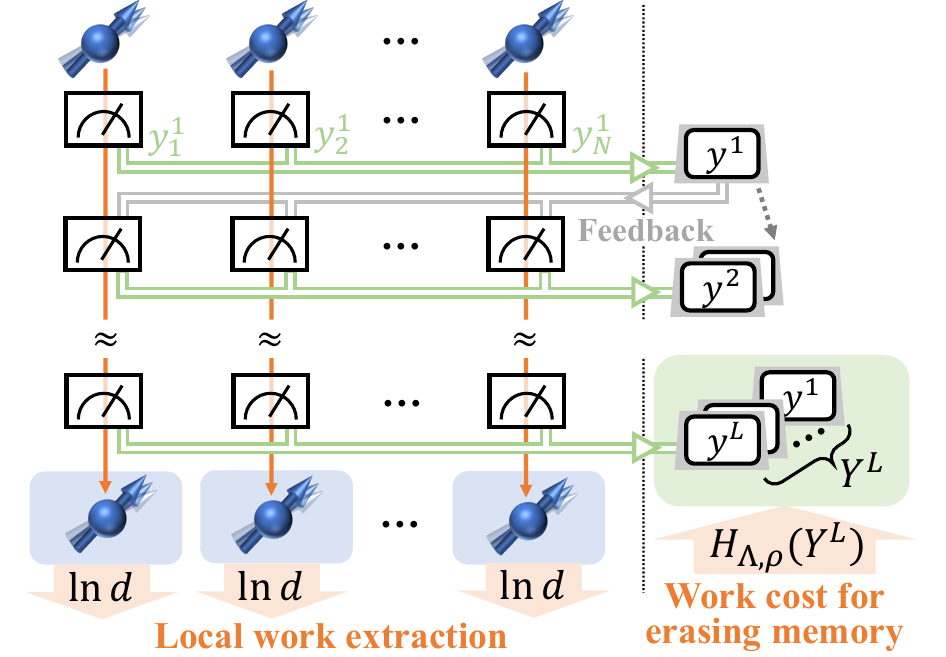}
    \caption{
    Schematic for an $L$-round LOCC protocol. 
    The local operation in the $l$-th round can be adaptively changed depending on the past measurement outcomes $Y^{l-1}$. 
    The net work extraction through the LOCC protocol consists of two contributions: 
    the extractable work from each local subsystem and the work cost of memory erasure.
    }
    \label{fig:LOCC}
\end{figure}

We next introduce the extractable work under LOCC, which constitutes an intermediate operational regime between global and strictly local operations.
As shown in Fig.~\ref{fig:LOCC}, a general LOCC protocol is composed of multiple rounds of local operations and classical communication, and includes adaptive operations conditioned on past measurement outcomes.
For an $L$-round LOCC protocol, let $y_n^l$ denote the measurement outcome at subsystem $n$ in round $l$, let $y^l=\{y_n^l\}_{n=1}^N$ be the set of outcomes obtained in round $l$, and let $Y^L=\{y^l\}_{l=1}^L$ be all outcomes obtained up to round $L$. It is not necessary to perform a nontrivial measurement on every subsystem in every round; the protocol may simply do nothing or apply a local unitary on a subsystem, in which case the corresponding outcome is assigned deterministically as $y_n^l = \phi$, representing a null outcome.

We can then formulate the extractable work by LOCC protocols with a fixed number of rounds $L$ as 
\begin{equation}
\label{eq:W_LOCC^L}
    W_{\rm LOCC}^L(\rho)
    = N \ln d - \min_{\Lambda \in \mathcal{M}^L} H_{\Lambda,\rho}(Y^L),
\end{equation}
where $H_{\Lambda,\rho}(Y^L)= - \sum_{Y^L} P_{\Lambda,\rho}(Y^L)\ln P_{\Lambda,\rho}(Y^L)$ is the Shannon entropy of the measurement outcomes, and $P_{\Lambda,\rho}(Y^L)$ is the probability of obtaining $Y^L$ when the protocol $\Lambda$ is applied. 
The set $\mathcal{M}^L$ contains only the $L$-round LOCC protocols whose final local measurements are rank-one, which suffices for the present purpose, as proved in the Supplemental Materials~\cite{SM}.
This expression for the extractable work can be understood by decomposing it into two contributions, as illustrated in Fig.~\ref{fig:LOCC}:
The first term, $N \ln d$, represents the sum of locally extractable work from post-measurement states, which is always pure since the local measurements in the final round are rank-one for $\Lambda\in \mathcal{M}^L$.
The second term, $H_{\Lambda,\rho}(Y^L)$, is the work required for erasing the classical memory that stores the measurement outcomes $Y^L$.
Based on Eq.~\eqref{eq:W_LOCC^L}, the extractable work under arbitrary LOCC protocols is given by
\begin{equation}
\label{eq:W_LOCC}
    W_{\rm LOCC}(\rho) = \sup_{L \in \mathbb{N}} W_{\rm LOCC}^{L}(\rho).
\end{equation}

We here note that this work-extraction framework has also been used to quantify correlations among subsystems.
Specifically, the difference between the extractable work under global operations and that under LOCC, $\Delta \equiv W_{\rm global}-W_{\rm LOCC},$ is called the \textit{work deficit}~\cite{oppenheim2002thermodynamical,horodecki2005local,zurek2003quantum} and is regarded as an operational measure of multipartite quantum correlations. Likewise, the gap between LOCC and strictly local operations, $\Delta_{\rm cl} \equiv W_{\rm LOCC}-W_{\rm local}$, is called the \textit{classical deficit}~\cite{synak2004classical} and quantifies classically accessible correlations.

\vspace{1mm}
\noindent
\textbf{Key observations:}
Based on the framework of work extraction, we now discuss the equivalence of pure states to the thermal state.
In the global and local regimes, the amount of extractable work is straightforward to quantify.
Specifically, $W_{\rm global} = N \ln d$ holds for any pure state, since it is determined solely by the global purity, as shown in Eq.~\eqref{eq:W_global}.
In contrast, the extractable work under local operations is determined by the reduced density operators, as shown in Eq.~\eqref{eq:W_local}.
In particular, when these reduced states are close to the infinite-temperature thermal state, i.e., the maximally mixed state, we obtain $W_{\rm local} = o(1)$, where $o(1)$ denotes a quantity that vanishes in the thermodynamic limit $N\to\infty$.
As summarized in the $W_{\rm local}$ column of Table~\ref{tab:summary}, such thermal behavior in the local regime has been established for various classes of many-body pure states, 
including Haar-random states~\cite{popescu2006entanglement,goldstein2006canonical,reimann2007typicality,tasaki2016typicality}, 
states sampled from approximate $t$-designs~\cite{brandao2016local,liu2018generalized}, 
graph states~\cite{briegel2001persistent,hein2006entanglement}, 
and states sampled from the subset state ensemble~\cite{aaronson2022quantum,giurgica2023pseudorandomness,jeronimo2023pseudorandom}.

The central question in this Letter is whether this equivalence between pure states and thermal states holds even under LOCC.
This is determined by the scaling of $W_{\rm LOCC}$: a state is regarded as equivalent to the thermal state when the extractable work is subextensive $W_{\rm LOCC} = o(N)$, and inequivalent when it is extensive $W_{\rm LOCC} = \Theta(N)$.
However, the exact evaluation of $W_{\rm LOCC}$ is very difficult in general, because Eqs.~\eqref{eq:W_LOCC^L} and \eqref{eq:W_LOCC} involve optimizations over all LOCC protocols, including arbitrarily many rounds of adaptive operations.
This makes a direct quantification of $W_{\rm LOCC}$ highly nontrivial.

A key idea to overcome this difficulty is to derive a relatively tractable upper bound on $W_{\rm LOCC}$:
\begin{equation}
\label{eq:W_LOCC_UB}
    W_{\rm LOCC}(\ket{\psi}) \leq N \ln d - E_g(\ket{\psi}),
\end{equation}
where $E_g(\ket{\psi}) \equiv -\ln \left( \max_{\ket{\bm{u}} \in \mathrm{PS}_N} |\braket{\bm{u}|\psi}|^2 \right)$
is a measure of multipartite entanglement called geometric entanglement~\cite{barnum2001monotones,wei2003geometric,shimony1995degree}, and $\mathrm{PS}_N$ denotes the set of $N$-qudit product states.
Although this quantity still involves an optimization over product states, it is far simpler than the optimization over arbitrary LOCC protocols appearing in $W_{\rm LOCC}$.
A derivation of Eq.~\eqref{eq:W_LOCC_UB} is provided in End Matter.

Equation~\eqref{eq:W_LOCC_UB} indicates that the equivalence to the thermal state under LOCC is governed by the amount of multipartite entanglement. 
Specifically, if the geometric entanglement is maximal up to subextensive correction, i.e., $E_g(\ket{\psi}) = N \ln d - o(N)$, then we have $W_{\rm LOCC} = o(N)$, implying the equivalence to the thermal state even under LOCC.
On the other hand, when the geometric entanglement is not asymptotically maximal, i.e., $E_g(\ket{\psi}) = N \ln d - \Theta(N)$, there may exist a protocol to achieve $W_{\rm LOCC} = \Theta(N)$, which means the inequivalence to the thermal state.

In this work, we address both of these cases.
First, we show the non-extensive work extraction for some classes of pure states by utilizing Eq.~\eqref{eq:W_LOCC_UB}.
Such classes of states include Haar-random states, states sampled from $t$-designs, and random graph states. 
Second, we demonstrate the extensive work extraction for the states with relatively simple multipartite entanglement structures, such as the constant-degree graph states and states sampled from the subset state ensemble. 
Since Eq.~\eqref{eq:W_LOCC_UB} only provides an upper bound on the extractable work, it cannot be used to establish extensive work extraction.
Therefore, we instead derive lower bounds on $W_{\rm LOCC}$ tailored to each state ensemble by explicitly constructing suitable work-extraction protocols.

\vspace{1mm}
\noindent
\textbf{Extractable work from highly entangled states:}
We here present the results for several classes of highly entangled states where the extractable work becomes non-extensive.
First, we address the pure states sampled from the uniform (Haar) distribution. 
In Ref.~\cite{gross2009most}, it was proved that the geometric entanglement $E_g$ of Haar-random states is asymptotically maximal with high probability.
The essence of the proof is to simplify the optimization appearing in the definition of $E_g$ by replacing the continuous set $\mathrm{PS}_N$ with a suitably chosen discrete set, denoted by $\mathcal{B}_N$. This reduction allows one to evaluate the overlaps $|\braket{\bm{v}|\psi}|$ only for a finite number of product states $\ket{\bm{v}} \in \mathcal{B}_N$.
By combining this result with Eq.~\eqref{eq:W_LOCC_UB}, we can obtain the following theorem.
\vspace{-2mm}
\begin{theorem}[\textbf{Haar-random states, informal}] \label{thm:Haar}
For a Haar-random $N$-qubit pure state $\ket{\psi}$, we have
\begin{equation}
\label{eq:Haar_W_LOCC}
     W_{\rm LOCC}(\ket{\psi}) \leq 2\ln N = o(N),
\end{equation}
with high probability.
\end{theorem}
\vspace{-1.5mm}
\noindent
This theorem implies that Haar-random pure states are equivalent to thermal states even when LOCC protocols are allowed.
The complete proof of this theorem is provided in the Supplemental Materials \cite{SM}.

We next consider pure states sampled from approximate state $t$-designs, whose thermal behavior has attracted considerable attention in recent years~\cite{kaneko2020characterizing,ho2022exact,ippoliti2022solvable,choi2023preparing,cotler2023emergent,ippoliti2023dynamical,mark2024maximum}.
An approximate state $t$-design is an ensemble of pure states whose $t$-th moment approximates that of the Haar distribution.
More precisely, an ensemble $\nu$ is called an $\varepsilon$-approximate state $t$-design if it satisfies the operator inequality
\begin{equation}
\label{eq:state_t_design}
    (1-\varepsilon)\,\Phi_{\rm Haar}^{(t)}
    \;\le\;
    \Phi_{\nu}^{(t)}
    \;\le\;
    (1+\varepsilon)\,\Phi_{\rm Haar}^{(t)},
\end{equation}
where $\Phi_{\nu}^{(t)} \equiv \mathbb{E}_{\psi\sim\nu}[\ket{\psi}\bra{\psi}^{\otimes t}]$ denotes the $t$-th moment operator, and $A\le B$ for Hermitian operators means that $B-A$ is positive semidefinite.
While several definitions of approximate state designs exist depending on how the deviation from the Haar distribution is quantified, we adopt the definition in Eq.~\eqref{eq:state_t_design} throughout this Letter.

For states sampled from such ensembles, we show that the geometric entanglement is asymptotically maximal, similarly to the case of Haar-random states.
The proof extends the strategy developed in Ref.~\cite{gross2009most} to state $t$-designs.
Specifically, we evaluate the overlaps $|\braket{\bm{v}|\psi}|$ for the elements $\ket{\bm{v}} \in \mathcal{B}_N$ by exploiting the $t$-design property, instead of using the property of Haar-random states.
By combining this result with Eq.~\eqref{eq:W_LOCC_UB}, we obtain the following theorem~\cite{SM}.
\vspace{-2mm}
\begin{theorem}[\textbf{Approximate $t$-designs, informal}]\label{thm:t_design}
Let $\nu$ be an $\varepsilon$-approximate state $t$-design on $N$-qudit pure states, with $\varepsilon=O(1)$ and $t=\omega(\ln N)$.  
Then, for $\ket{\psi}\sim\nu$, we have
\begin{equation}
\label{eq:W_LOCC_t_design}
    W_{\rm LOCC}(\ket{\psi})=o(N),
\end{equation}
with high probability.
\end{theorem}
\vspace{-1.5mm}
\noindent
This theorem implies that states sampled from approximate state $t$-designs with $t=\omega(\ln N)$ are also equivalent to thermal states under LOCC, since their extractable work remains non-extensive.

Since state $t$-designs concern moments only up to order $t$, Theorem~\ref{thm:t_design} includes broader situations than Theorem~\ref{thm:Haar}. 
In particular, $t$-design ensembles are known to be generated much more efficiently compared to Haar-random states \cite{brandao2016local,haferkamp2022random,chen2024incompressibility,schuster2024random,laracuente2024approximate,yada2025non}, implying that Theorem~\ref{thm:t_design} addresses a more experimentally practical setting. Specifically, a state $t$-design can be generated with $O(t\,\polylog t \cdot \ln N)$-depth random circuit \cite{schuster2024random,laracuente2024approximate}, whereas Haar-random states require more than $e^{\Omega(N)}$-depth circuit.

Another example of highly entangled states is random graph states, which have been widely studied as analytically tractable models of quantum many-body systems~\cite{zhou2022entanglement,ghosh2023complexity,GhoshPRXQ2025}.
For a graph $G=(V,E)$ with vertex set $V$ and edge set $E$, the corresponding graph state is defined as
\begin{equation}
    \ket{G} \equiv \left( \prod_{(i,j)\in E} CZ_{ij} \right) \bigotimes_{n \in V}\ket{+}_n,
\end{equation}
where $CZ_{ij}$ is the controlled-$Z$ gate acting on qubits $i$ and $j$, and $\ket{+}_n\equiv (\ket{0}_n+\ket{1}_n)/\sqrt{2}$.
Such a state generalizes the cluster state to arbitrary graph connectivity, which serves as a universal resource state for measurement-based quantum computation~\cite{briegel2001persistent}.
When the graph is sampled uniformly at random from all $N$-vertex graphs, the degree of a typical vertex scales as $r \sim N/2$, indicating a highly complex multipartite entanglement structure.

For such random graph states, Ref.~\cite{GhoshPRXQ2025} showed that the geometric entanglement is asymptotically maximal with high probability.
By combining this result with Eq.~\eqref{eq:W_LOCC_UB}, we obtain the following theorem~\cite{SM}.
\vspace{-2mm}
\begin{theorem}[\textbf{Random graph states, informal}]\label{thm:graph}
For a graph state $\ket{G}$ sampled from the uniform distribution over $N$-vertex graphs,
\begin{equation}
    W_{\rm LOCC}(\ket{G}) = o(N),
\end{equation}
with high probability.
\end{theorem}
\vspace{-1.5mm}
\noindent
This theorem thus implies that random graph states are equivalent to thermal states even when LOCC protocols are allowed.

\vspace{1mm}
\noindent
\textbf{Extractable work from constant-degree graph states:}
On the other hand, for states with relatively simple entanglement structures, the extractable work under LOCC can become extensive.
As a first example, we consider constant-degree graph states, namely graph states in which the degree of every vertex is fixed to a constant $r = O(1)$.
Typical examples include the hexagonal, square, and triangular lattices, which have fixed degrees $r = 3$, $4$, and $6$, respectively.
Since the degree $r$ characterizes the complexity of a graph state~\cite{ghosh2023complexity,GhoshPRXQ2025}, these states are significantly less complex than the random graph states considered in Theorem~\ref{thm:graph}, whose degree scales extensively as $r \sim N/2$.

For any constant-degree graph state, the extractable work under strictly local operations vanishes, because the reduced density operator of each qubit is always maximally mixed.
In contrast, when LOCC protocols are allowed, we show that the extractable work satisfies
\begin{equation}
\label{eq:d_regular_graph_W_LOCC}
    W_{\rm LOCC}(\ket{G}) \ge \frac{N\ln 2}{r+1} = \Theta(N).
\end{equation}
This bound is derived by explicitly constructing a work-extraction protocol, which is described in detail in the Supplemental Materials \cite{SM}.
Equation~\eqref{eq:d_regular_graph_W_LOCC} therefore shows that constant-degree graph states are not equivalent to the thermal state under LOCC,
in sharp contrast to random graph states, whose multipartite entanglement is asymptotically maximal.

\vspace{1mm}
\noindent
\textbf{Extractable work from subset state ensemble:}
As another example of states with a relatively simple entanglement structure, we consider pure states that exhibit limited coherence in the computational basis.
Specifically, we focus on the subset state ensemble~\cite{aaronson2022quantum,giurgica2023pseudorandomness,jeronimo2023pseudorandom} on an $N$-qubit system, defined as
\begin{equation}
\label{eq:subset_state_ens}
    \nu_K \equiv \left\{
    \ket{\psi_S}\equiv \frac{1}{\sqrt{K}}\sum_{z\in S} \ket{z}
    \,\middle|\,
    S\subset\{0,1\}^N,\ |S|=K
    \right\},
\end{equation}
where $K$ denotes the support size of the superposition.
Each state $\ket{\psi_S}$ is a uniform superposition over the $K$ computational basis states contained in the subset $S$, and the ensemble $\nu_K$ consists of all such states with fixed $|S|=K$.
It has been shown that a state sampled from $\nu_K$ is locally thermalized, i.e., $W_{\rm local}=o(1)$, provided that the support size satisfies
$e^{\omega(\ln N)} < K < 2^{cN}$ for some constant $c<1$~\cite{giurgica2023pseudorandomness,jeronimo2023pseudorandom}.

In contrast, when LOCC protocols are allowed, we can show that the extractable work is lower bounded as
\begin{equation}
\label{eq:subset_W_LOCC}
    W_{\rm LOCC}(\ket{\psi_S}) \ge N\ln 2 - \ln K = \Theta(N).
\end{equation}
This bound is achieved by a simple one-round protocol $\Lambda$ that performs projective measurements in the computational basis $\{\ket{0},\ket{1}\}$ on every qubit.
Since the state $\ket{\psi_S}$ has support size $K$, the Shannon entropy of the measurement outcomes is $H_{\Lambda,\ket{\psi_S}}(y^1)=\ln K$, which directly leads to Eq.~\eqref{eq:subset_W_LOCC}.
This result shows that states sampled from the subset state ensemble are inequivalent to the thermal state under LOCC, despite being locally indistinguishable from thermal states.

Importantly, this conclusion is not restricted to the subset state ensemble itself, but applies more broadly to pure states with low coherence in the computational basis.
In particular, Eq.~\eqref{eq:subset_W_LOCC} holds for any pure state whose support size satisfies $K < 2^{cN}$ for some constant $c<1$, implying inequivalence to the thermal state under LOCC.
A notable example is provided by the output states of automaton circuits~\cite{fisher2023random,iaconis2019anomalous,iaconis2021quantum,iaconis2020measurement}, 
which are circuits that do not generate coherence with respect to the computational basis.
In particular, recent studies have identified classes of automaton circuits whose output states exhibit local thermalization while retaining low coherence~\cite{gopalakrishnan2018facilitated,gopalakrishnan2018operator,klobas2021exact,klobas2021exact_rel,klobas2021entanglement,alba2019operator,feng2024dynamics,bertini2025quantum,szasz2025entanglement}.
These states are therefore equivalent to the thermal state under strictly local operations, but inequivalent under LOCC, in the same way as the subset state ensemble.

\vspace{1mm}
\noindent
\textbf{Summary and outlook:} 
In this work, we have investigated which classes of pure states are equivalent to thermal states under LOCC, based on the scaling of the extractable work. 
To this end, we first established Eq.~\eqref{eq:W_LOCC_UB}, which implies that states with asymptotically maximal multipartite entanglement cannot yield extensive work under LOCC, and are therefore equivalent to thermal states. 
We further showed that this class of highly entangled states includes Haar-random states, approximate state $t$-designs with superlogarithmic $t$, and random graph states, leading to Theorems~\ref{thm:Haar}, \ref{thm:t_design}, and \ref{thm:graph}. 
In contrast, we demonstrated extensive work extraction for states with limited multipartite entanglement, such as constant-degree graph states and states sampled from the subset state ensemble.
Thus, our work extends the understanding of thermal behavior in isolated quantum many-body systems beyond the traditional local regime, a setting that is becoming increasingly accessible in modern experiments. 

Finally, we present several future research directions.
An important open question is whether there exist classes of many-body quantum states whose geometric entanglement is not asymptotically maximal, yet whose extractable work under LOCC remains subextensive. Addressing this question would require developing upper bounds different from those in Eq.~\eqref{eq:W_LOCC_UB}.
It is also intriguing to quantify the extractable work under LOCC in real quantum experiments, thereby testing the equivalence to thermal states.
Since the extractable work is defined in terms of the Shannon entropy of local measurement outcomes as shown in Eq.~\eqref{eq:W_LOCC^L}, we expect that it can be estimated in a manner similar to those used in random circuit sampling experiments~\cite{choi2023preparing,arute2019quantum,morvan2024phase,zhu2022quantum,gao2025establishing}, in which outcome statistics are obtained from repeated measurements.
While the goal of random circuit sampling is to benchmark computational complexity, estimating the extractable work would provide an operational probe for the structure of multipartite quantum correlations.

\vspace{3mm}

\begin{acknowledgments}
We thank Shoki Sugimoto for fruitful discussions.
T.Y. is supported by World-leading Innovative Graduate Study Program for Materials Research, Information, and Technology (MERIT-WINGS) of the University of Tokyo. T.Y. is also supported by JSPS KAKENHI Grant No. JP23KJ0672.
T.S. and N.Y. are supported by JST ERATO Grant Number JPMJER2302, Japan, and by Institute of AI and Beyond of the University of Tokyo.
T.S. is also supported by JST CREST Grant Number JPMJCR20C1.
N.Y. wishes to thank JST PRESTO No. JPMJPR2119, JST Grant Number JPMJPF2221, JST CREST Grant Number JPMJCR23I4, and IBM Quantum.
\end{acknowledgments}

\bibliography{bibliography}

\clearpage

\appendix

\section*{End Matter}
\section{Derivation of Eq.~\eqref{eq:W_LOCC_UB}} \label{app:derive_UB}

In this appendix, we derive the inequality
\begin{equation}
     W_{\rm LOCC}(\ket{\psi}) \leq N \ln d - E_g(\ket{\psi}), \tag{\ref{eq:W_LOCC_UB}}
\end{equation}
where $E_g(\ket{\psi}) \equiv -\ln \left( \max_{\ket{\bm{u}} \in \mathrm{PS}_N} |\braket{\bm{u}|\psi}|^2 \right)$
is the geometric entanglement and $\mathrm{PS}_N$ is the set of $N$-qudit product states.

To this end, we first introduce the description of an $L$-round LOCC protocol $\Lambda \in \mathcal{M}^L$.
The set of Kraus operators for the local measurement on the $n$-th subsystem in the $l$-th round is denoted by 
$\{K_{y_n^{l}}^{Y^{l-1}}\}_{y_n^{l}}$, where the superscript $Y^{l-1}$ indicates that this measurement may depend on the previous outcomes $Y^{l-1}$.
The Kraus operators for the entire $L$-round protocol are denoted by $\{\mathcal{K}_{Y^{L}}\}_{Y^{L}}$ and are given by
\begin{equation}
\label{eq:Kraus_L_LOCC}
    \mathcal{K}_{Y^{L}}
    \equiv 
    \prod_{l=1}^L \left(\bigotimes_{n=1}^N K_{y_n^{l}}^{Y^{l-1}}\right)
    = 
    \bigotimes_{n=1}^N \left(\prod_{l=1}^L K_{y_n^{l}}^{Y^{l-1}}\right),
\end{equation}
where the first expression emphasizes that the measurements are performed successively for $L$ rounds, and the second expression makes explicit that the overall operator is a tensor product over the $N$ subsystems.

For any protocol $\Lambda \in \mathcal{M}^L$, the local measurements in the final round are rank-one.
Using this property, the Kraus operator for outcome $Y^L$ can be expressed as
\begin{equation}
\label{eq:Kraus_L_LOCC_rank_one}
    \mathcal{K}_{Y^L}
    = 
    \alpha_{Y^L}
    \bigotimes_{n=1}^N\,
    \ket{\gamma_{Y^L}}_{\!n}\raisebox{-0.4ex}{${}_n$}\!\!\bra{\beta_{Y^L}},
\end{equation}
where $\alpha_{Y^L}$ is the nonnegative parameter satisfying $0 \leq \alpha_{Y^L} \leq 1$, and $\ket{\beta_{Y^L}}_n$ and $\ket{\gamma_{Y^L}}_n$ are pure states on the $n$-th subsystem, both determined by $Y^L$.

Using this representation, we now derive Eq.~\eqref{eq:W_LOCC_UB}:
\begin{align}
W_{\rm LOCC}^{L}(\ket{\psi})
    &= N \ln d - \min_{\Lambda \in \mathcal{M}^L} H_{\Lambda,\ket{\psi}}(Y^L) \nonumber\\
    &\leq N \ln d - \min_{\Lambda \in \mathcal{M}^L}
        \left[-\ln\!\left(\max_{Y^L} P_{\Lambda,\ket{\psi}}[Y^L]\right)\right] 
        \nonumber\\
    &\leq N \ln d - E_g(\ket{\psi}). 
    \label{eq:W_LOCC^L_UB}
\end{align}
Here, the second line follows from the fact that the Shannon entropy of the distribution 
$\{P_{\Lambda,\ket{\psi}}[Y^L]\}_{Y^L}$ is lower bounded by its min-entropy.
The final inequality is obtained from the relation
\begin{align*}
    P_{\Lambda,\ket{\psi}}[Y^L]
    &= 
    \alpha_{Y^L}^2 
    \left| \left(\bigotimes_{n=1}^N \raisebox{-0.4ex}{${}_n$}\!\!\bra{\beta_{Y^L}} \right) \ket{\psi} \right|^2 \\
    &\leq 
    \max_{\ket{\bm{u}} \in \mathrm{PS}_N}
    \left| \braket{\bm{u}|\psi} \right|^2,
\end{align*}
where the first line follows from Eq.~\eqref{eq:Kraus_L_LOCC_rank_one}, and the second line follows from $0\leq \alpha_{Y^L} \leq 1$.
Since Eq.~\eqref{eq:W_LOCC^L_UB} holds for arbitrary $L$, the supremum over $L$ is also bounded by the same quantity, which completes the derivation.

\end{document}


\title{Supplemental Material for ``Testing the equivalence to thermal states via extractable work under LOCC''}

\author{Toshihiro Yada}
\affiliation{\mbox{Department of Applied Physics, The University of Tokyo, 7-3-1 Hongo, Bunkyo-ku, Tokyo 113-8656, Japan}}

\author{Nobuyuki Yoshioka}
\affiliation{\mbox{International Center for Elementary Particle Physics, The University of Tokyo, 7-3-1 Hongo, Bunkyo-ku, Tokyo 113-0033, Japan}}

\author{Takahiro Sagawa}
\affiliation{\mbox{Department of Applied Physics, The University of Tokyo, 7-3-1 Hongo, Bunkyo-ku, Tokyo 113-8656, Japan}}

\setcounter{section}{0}
\setcounter{equation}{0}
\setcounter{figure}{0}
\setcounter{table}{0}
\setcounter{page}{1}
\renewcommand{\thesection}{S\arabic{section}}
\renewcommand{\theequation}{S\arabic{equation}}
\renewcommand{\thefigure}{S\arabic{figure}}
\renewcommand{\thetable}{S\arabic{table}}
\renewcommand{\bibnumfmt}[1]{[S#1]}
\renewcommand{\citenumfont}[1]{S#1}

\maketitle

\tableofcontents

\vspace{4mm}\noindent

\section{Formulation of extractable work under LOCC}
In this section, we formulate the extractable work of an $N$-qudit state under LOCC (local operations and classical communication). 
While the extractable work under LOCC has been formulated in previous works~\cite{oppenheim2002thermodynamical,horodecki2005local}, 
these frameworks mainly focus on settings in which only projective measurements are allowed on each local subsystem.
Here, we extend the formulation to allow arbitrary POVMs (positive operator-valued measures) on local subsystems, thereby encompassing general LOCC protocols.

In Sec.~\ref{ss:desc_LOCC}, we introduce LOCC protocols and define the extractable work by a fixed protocol.
In Sec.~\ref{ss:exp_LOCC}, we derive explicit expressions for the extractable work, which are presented in Eqs.~(3) and (4) of the main text.

\subsection{Description of LOCC protocols}\label{ss:desc_LOCC}

We here introduce the description of LOCC protocols.
In general, an LOCC protocol is composed of multiple rounds of local operations and classical communication \cite{chitambar2014everything}.
We now consider an $L$-round protocol $\Lambda$. 
The measurement outcome obtained from the $n$-th subsystem in the $l$-th round is denoted by $y_{n}^l$, the collection of outcomes from all the subsystems in the $l$-th round is denoted by $y^l \equiv \{y_{n}^l\}_{n=1}^{N}$, and the set of all the outcomes until the $L$-th round is denoted by $Y^{L} \equiv \{y^l\}_{l=1}^{L}$. 
Here, the measurement in the $l$-th round can depend on the past measurement outcomes $Y^{l-1}$.
The set of Kraus operators for subsystem $n$ in round $l$ is denoted by $\{K_{y_n^l}^{Y^{l-1}}\}_{y_n^l}$. 
The measurement on each subsystem can be an arbitrary POVM, and the only requirement is that it satisfies the completeness condition $\sum_{y_{n}^l} K_{y_{n}^l}^{Y^{l-1}\dag} K_{y_{n}^l}^{Y^{l-1}} = \mathbbm{1}_n$ for all $l$, $n$, and $Y^{l-1}$, where $\mathbbm{1}_n$ represents the identity operator on the $n$-th subsystem. 

The Kraus operators for the entire $L$-round protocol are denoted by $\{\mathcal{K}_{Y^{L}}\}_{Y^{L}}$, and they are given by
\begin{equation}
\label{eq:Kraus_L_LOCC}
    \mathcal{K}_{Y^{L}}
    \equiv 
    \prod_{l=1}^L \left(\bigotimes_{n=1}^N K_{y_n^{l}}^{Y^{l-1}}\right)
    = 
    \bigotimes_{n=1}^N \left(\prod_{l=1}^L K_{y_n^{l}}^{Y^{l-1}}\right).
\end{equation}
By using these Kraus operators, the probability of obtaining the measurement outcome $Y^{l}$, denoted by $P_{\Lambda,\rho}(Y^{l})$, and the conditional density operator right after the $l$-th round, denoted by $\rho^{l,Y^l}$, satisfy $P_{\Lambda,\rho} (Y^{l})\rho^{l,Y^l} = \mathcal{K}_{Y^{l}} \rho \mathcal{K}_{Y^{l}}^\dag$.

Next, we define the extractable work with a fixed $L$-round LOCC protocol, denoted by $\Lambda$.
The extractable work from $\rho$ through the protocol $\Lambda$ is given by
\begin{equation}
\label{eq:work_lambda_adap}
    W_\Lambda (\rho) \equiv \sum_{n=1}^N \left\{\ln d - \sum_{Y^{L}}P_{\Lambda,\rho}(Y^{L})S(\rho_{n}^{L,Y^{L}})\right\} - H_{\Lambda,\rho} (Y^{L}),
\end{equation}
where $H_{\Lambda,\rho} (Y^{L}) \equiv -\sum_{Y^{L} }P_{\Lambda,\rho} (Y^{L})\ln P_{\Lambda,\rho} (Y^{L})$ denotes the Shannon entropy for the measurement outcome $Y^L$, and $\rho_{n}^{L,Y^{L}} \equiv \tr_{\overline{n}}[\rho^{L,Y^L}]$ is the reduced density operator on the $n$-th subsystem.
Equation~\eqref{eq:work_lambda_adap} is decomposed into two contributions:  
The first term in the curly brackets represents the extractable work from each local subsystem after the protocol $\Lambda$. This follows from the fact that when the measurement outcomes $Y^L$ are obtained, the extractable work from the $n$-th subsystem is given by $\ln d - S(\rho_{n}^{L,Y^{L}})$.  
The second contribution is the Shannon entropy term, which represents the work cost required for the memory erasure. Since we can gather all memory registers through classical communication and erase them collectively, the work cost required for this erasure is given by the Shannon entropy $H_{\Lambda,\rho} (Y^{L})$.

\subsection{Derivation of the extractable work under LOCC} \label{ss:exp_LOCC}

We now derive the expression for the extractable work under LOCC, which is given by Eqs.~(3) and (4) in the main text.
In the derivation, the following lemma plays a central role.
\vspace{-2mm}
\begin{lemma}\label{lem:formulation_two_way_LOCC}
Let $\mathcal{N}^{L}$ be the set of arbitrary $L$-round LOCC protocols, and let $\mathcal{M}^{L}$ be the set of $L$-round LOCC protocols whose local measurements in the $L$-th (final) round are all rank-one.
Then, for any protocol $\Lambda \in \mathcal{N}^L$, there exists a protocol $\Lambda^\prime \in \mathcal{M}^{L+1}$ such that
\begin{equation}
\label{eq:W_Lambda_M^L}
    W_{\Lambda} (\rho) \leq W_{\Lambda^\prime}(\rho).
\end{equation}
\end{lemma}

\noindent
Using this lemma, we derive the expressions given in Eqs.~(3) and (4) of the main text.
We first introduce the following notations:
\begin{align}
\widetilde{W}_{\rm LOCC}^{L}(\rho) 
&\equiv \max_{\Lambda \in  \mathcal{N}^{L}} W_{\Lambda}(\rho), 
\label{eq:def_TildeW^L}\\
W_{\rm LOCC}^{L}(\rho) 
&\equiv \max_{\Lambda \in  \mathcal{M}^{L}} W_{\Lambda}(\rho) 
= N \ln d - \min_{\Lambda \in \mathcal{M}^{L}} H_{\Lambda,\rho}(Y^L), 
\label{eq:def_W^L}
\end{align}
where the second equality in Eq.~\eqref{eq:def_W^L} follows from the fact that, for $\Lambda \in \mathcal{M}^{L}$, 
the conditional reduced density operator is pure, i.e., $S(\rho_{n}^{L,Y^{L}}) = 0$ for all $n$ and $Y^{L}$, 
because the final-round measurements are all rank-one.
For these quantities, we have the inequality
\begin{equation}
\label{eq:W^L_TildeW^L}
    W_{\rm LOCC}^{L}(\rho) 
    \leq \widetilde{W}_{\rm LOCC}^{L}(\rho)  
    \leq W_{\rm LOCC}^{L+1}(\rho),
\end{equation}
where the first inequality follows from the inclusion relation $\mathcal{M}^{L} \subset \mathcal{N}^{L}$, 
and the second inequality follows from Lemma~\ref{lem:formulation_two_way_LOCC}.

Therefore, the extractable work under arbitrary LOCC protocols is given by
\begin{align}
    W_{\rm LOCC} (\rho) 
    &\equiv  \sup_{L \in\mathbb{N}} \widetilde{W}_{\rm LOCC}^{L}(\rho) 
    \nonumber \\
    &= \sup_{L \in\mathbb{N}} W_{\rm LOCC}^{L}(\rho),
\label{eq:W_LOCC}
\end{align}
where the number of rounds $L$ is also optimized.
Here, the suprema in both lines are well-defined since the sequences 
$\{\widetilde{W}_{\rm LOCC}^{L}(\rho)\}_{L \in\mathbb{N}}$ 
and $\{W_{\rm LOCC}^{L}(\rho)\}_{L \in\mathbb{N}}$ 
are monotonically increasing in $L$ and are bounded above by $N \ln d$.
Furthermore, Eq.~\eqref{eq:W^L_TildeW^L} implies that these two suprema take the same value, 
which yields the second line of Eq.~\eqref{eq:W_LOCC}.

\medskip
\noindent
\textit{Proof of Lemma~\ref{lem:formulation_two_way_LOCC}.}
We prove Eq.~\eqref{eq:W_Lambda_M^L} by explicitly constructing a protocol 
$\Lambda^\prime \in \mathcal{M}^{L+1}$ corresponding to a given $\Lambda \in \mathcal{N}^L$.
Up to the $L$-th round, the operations of $\Lambda^\prime$ are chosen to be identical to those of $\Lambda$.
In the $(L+1)$-th round, the local measurement on the $n$-th subsystem is determined by the conditional reduced density operator 
$\rho_{n}^{L,Y^{L}}$, whose eigenvalues and eigenstates are denoted by 
$\{q_n^{Y^{L}}(x)\}_{x}$ and $\{\ket{x}^{\,Y^{L}}_{n}\}_{x}$, respectively, such that
\[
\rho_{n}^{L,Y^{L}} 
= \sum_{x} q_n^{Y^{L}}(x) \,
\ket{x}^{\,Y^{L}}_{n} {}^{\,Y^{L}}_{n}\!\bra{x}.
\]
Specifically, in the $(L+1)$-th round of $\Lambda^\prime$, 
rank-one projective measurements in the basis $\{\ket{x}^{\,Y^{L}}_{n}\}_{x}$ 
are performed on subsystem $n$, conditioned on the past outcomes $Y^{L}$.
The probability of obtaining outcomes $Y^{L+1}$ is then given by
\begin{equation}
\label{eq:prob_Lambda^prime}
    P_{\Lambda^\prime,\rho}(Y^{L+1}) 
    = P_{\Lambda,\rho}(Y^L) 
    \prod_{n=1}^N  q_n^{Y^{L}}(y_n^{L+1}).
\end{equation}

For the protocols $\Lambda$ and $\Lambda^\prime$, we obtain
\begin{equation}
\begin{split}
    W_{\Lambda}(\rho) 
    &\equiv \sum_{n=1}^N  
    \left\{ 
        \ln d - \sum_{Y^{L}} P_{\Lambda,\rho}(Y^{L}) 
        S(\rho_{n}^{L,Y^{L}}) 
    \right\}  
    - H_{\Lambda,\rho}(Y^L) \\
    &= N \ln d 
    - \sum_{n=1}^N H_{\Lambda^\prime,\rho}(y_n^{L+1} \mid Y^{L}) 
    - H_{\Lambda^\prime,\rho}(Y^L)\\
    &\leq N\ln d 
    - \sum_{n=1}^N 
    H_{\Lambda^\prime,\rho}\!\left(
        y_n^{L+1} \mid Y^{L},\{y_i^{L+1}\}_{i=1}^{n-1}
    \right)
    - H_{\Lambda^\prime,\rho}(Y^L)\\
    &= N\ln d - H_{\Lambda^\prime,\rho}(Y^{L+1}) \\
    &\equiv W_{\Lambda^\prime}(\rho),
\end{split}
\end{equation}
where the conditional Shannon entropy is defined as
$H(x|y)\equiv-\sum_{x,y} p(x,y) \ln [p(x,y)/p(y)]$.
The second line follows from Eq.~\eqref{eq:prob_Lambda^prime}, 
the third line follows from the fact that conditioning does not increase the Shannon entropy, 
and the fourth line follows from the chain rule for the Shannon entropy.
\qed

\section{Extractable work from Haar-random pure states under LOCC}

In this section, we provide the complete proof of Theorem~1 in the main text. 
\begin{theorem}[Theorem~1 in the main text]\label{thm:Haar_2LOCC}
Let $\mu_{\rm H}$ be the uniform distribution over $N$-qudit pure states. Then the following inequality holds:
\begin{equation}
\label{eq:work_2LOCC}
    \ProbH \left[ W_{\rm LOCC} (\ket{\psi}) \geq  2 \ln N \right] \leq e^{- O(N^2)}.
\end{equation}
\end{theorem}

The proof of this theorem consists of two steps. 
In the first step, we derive an upper bound on $W_{\rm LOCC}$,
\begin{equation}
\label{eq:W_LOCC_UB}
    W_{\rm LOCC}(\ket{\psi}) \leq N \ln d - E_g(\ket{\psi}),
\end{equation}
where $E_g(\ket{\psi}) \equiv -\ln \!\left(\max_{\ket{\bm{u}} \in \mathrm{PS}_N} |\bra{\bm{u}} \psi \rangle |^2 \right)$ is the geometric entanglement and $\mathrm{PS}_N$ is the set of the $N$-qudit product states.
This upper bound is given in Eq.~(5) of the main text, and its derivation is provided in Appendix A.
In the second step, we show that this upper bound satisfies $N \ln d - E_g(\ket{\psi}) \leq 2 \ln N$ with high probability. The main technical challenge in this step is to estimate the value of the geometric entanglement $E_g$, which involves the optimization over $\mathrm{PS}_N$. To handle this optimization, we use Lemmas~\ref{lem:e_net_g_ent} and~\ref{lem:concent_Haar}, derived in this section.
We note that the essential proof idea of the second step follows the same strategy as that used in Theorem~1 of Ref.~\cite{gross2009most}, 
where it was shown that the geometric entanglement is asymptotically maximal for Haar-random states. 
Nevertheless, we present the proof here explicitly, since Lemma~\ref{lem:e_net_g_ent} involves a technical improvement that is crucial for its application to state $t$-designs in Sec.~\ref{s:t_design}.

In Sec.~\ref{ss:approx_PS}, we derive Lemma~\ref{lem:e_net_g_ent}, which introduces a discrete set of product states that well approximates $\mathrm{PS}_N$. In Sec.~\ref{ss:tail_bound_Haar}, we prove Lemma~\ref{lem:concent_Haar}, which shows that the inner product between a Haar-random state $\ket{\psi}$ and a fixed product state $\ket{\bm{v}}$ satisfies $-\ln |\bra{\bm{v}} \psi \rangle|^2 = N \ln d - o(N)$ with high probability. Finally, in Sec.~\ref{ss:prf_thm1}, we complete the proof of Theorem~\ref{thm:Haar_2LOCC} by combining these lemmas.

\subsection{Approximation of the set of product states} \label{ss:approx_PS}

In this subsection, we prove that there exists a discrete set of $N$-qudit product states with sufficiently small cardinality that well approximates $\mathrm{PS}_N$. Specifically, we establish the following lemma:
\begin{lemma} \label{lem:e_net_g_ent}
There exists a discrete set of $N$-qudit product states $\mathcal{B}_N$ whose cardinality is upper bounded as $|\mathcal{B}_N|\leq e^{2d N (\ln N +3)}$, and which satisfies the following inequality for any $N$-qudit pure state $\ket{\psi}$:
\begin{equation}
\label{seq:e_net_g_ent}
     \max_{\ket{\bm{u}} \in \mathrm{PS}_N} |\bra{\bm{u}} \psi \rangle |
     \leq
     2 \cdot \left\{ \max_{\ket{\bm{v}} \in \mathcal{B}_N} |\bra{\bm{v}} \psi \rangle | \right\}.
\end{equation}
\end{lemma}
\noindent
While this lemma has already been proved, e.g., in Ref.~\cite{GhoshPRXQ2025}, we include the proof here to make the discussion self-contained.

To prove this lemma, we use the notion of an $\varepsilon$-net. A set of $d$-dimensional pure states $\mathcal{A}$ is said to be an $\varepsilon$-net if, for any pure state $\ket{\psi}$, there exists an element $\ket{v} \in \mathcal{A}$ satisfying $\|\ket{\psi} - \ket{v}\| \leq \varepsilon$, where $\|\cdot\|$ denotes the Euclidean norm. In other words, an $\varepsilon$-net is a discrete set that covers all pure states within an error of $\varepsilon$.
In previous work~\cite{hayden2004randomizing}, the existence of an $\varepsilon$-net of pure states with an upper bound on its cardinality was proved as follows:
\begin{lemma}\label{lem:UB_cardinality_e-net}
There exists an $\varepsilon$-net of $d$-dimensional pure states $\mathcal{A}_{\varepsilon}$ whose cardinality is upper bounded as 
\begin{equation*}
    |\mathcal{A}_{\varepsilon}| \leq \left( \frac{5}{\varepsilon} \right)^{2d}.
\end{equation*}
\end{lemma}

By utilizing Lemma~\ref{lem:UB_cardinality_e-net}, we can derive Lemma~\ref{lem:e_net_g_ent} as follows:
\begin{proof}[Proof of Lemma~\ref{lem:e_net_g_ent}]
First, we explicitly introduce a discrete set $\mathcal{B}_N$ as 
\begin{equation*}
    \mathcal{B}_N \equiv \left\{ \ket{\bm{v}} \equiv \bigotimes_{i=1}^N \ket{v_i}\ \middle |\  \ket{v_i} \in \mathcal{A}_{\ln(3/2)/N} \right\},
\end{equation*}
where $\mathcal{A}_{\varepsilon}$ represents an $\varepsilon$-net for the $d$-dimensional pure states of each local subsystem.
From this construction and Lemma~\ref{lem:UB_cardinality_e-net}, we can immediately obtain the upper bound on the cardinality of $\mathcal{B}_N$ as
\begin{equation}
    |\mathcal{B}_N| = | \mathcal{A}_{\ln(3/2)/N}|^N \leq \left(\frac{5N}{\ln(3/2)}\right)^{2dN} \leq e^{2d N (\ln N +3)}. 
\end{equation}

Now, what remains is to prove Eq.~\eqref{seq:e_net_g_ent} with this discrete set $\mathcal{B}_N$.
We define the product state that achieves the maximum value on the left-hand side of Eq.~\eqref{seq:e_net_g_ent} as $\ket{\bm{u}^*}\equiv \bigotimes_{i=1}^N \ket{u_i^*}$. Then, each $d$-dimensional pure state $\ket{u_i^*}$ can be represented as $\ket{u_i^*} = \ket{v_i^0} + \delta_i \ket{v_i^1}$ with an element $\ket{v_i^0} \in \mathcal{A}_{\ln(3/2)/N}$ and a normalized pure state $\ket{v_i^1}$. By choosing an appropriate $\ket{v_i^0}$, $\delta_i$ can always be taken as $\delta_i \leq \frac{\ln(3/2)}{N}$ since $\mathcal{A}_{\ln(3/2)/N}$ is an $(\ln(3/2)/N)$-net of the $d$-dimensional pure states.
Based on these representations, we can derive the following inequality:
\begin{align}
\max_{\ket{\bm{u}} \in \mathrm{PS}_N} |\langle \bm{u}|\psi \rangle |  &= \left| \left(\bigotimes_{i=1}^N \bra{v_i^{0}} + \delta_i \bra{v_i^{1}}\right) \ket{\psi}\right| \nonumber\\
&\leq \sum_{j_1,j_2,\dots,j_N=\{0,1\}} \left[\prod_{i=1}^N (\delta_i)^{j_i}\right] \cdot\left| \left(\bigotimes_{i=1}^N \bra{v_i^{j_i}} \right) \ket{\psi}\right| \nonumber\\
&\leq \left|\left(\bigotimes_{i=1}^N \bra{v_i^{0}} \right) |\psi \rangle \right| + \sum_{n=1}^N \left(\begin{array}{c}
        N \\
        n
    \end{array}\right) \cdot\left(\frac{\ln(3/2)}{N}\right)^n \cdot \max_{\ket{\bm{u}} \in \mathrm{PS}_N} |\bra{\bm{u}} \psi \rangle | \nonumber\\
&= |\langle \bm{v}^0|\psi \rangle | + \left[\left(1+\frac{\ln(3/2)}{N}\right)^N-1\right]\cdot \max_{\ket{\bm{u}} \in \mathrm{PS}_N} |\bra{\bm{u}} \psi \rangle| \nonumber \\
&\leq  |\langle \bm{v}^0|\psi \rangle | + \frac{1}{2} \max_{\ket{\bm{u}} \in \mathrm{PS}_N} |\bra{\bm{u}} \psi \rangle|, \label{eq:max_prod_ent_g}
\end{align}
where we use the notation $\ket{\bm{v}^0} \equiv \bigotimes_{i=1}^N \ket{v_i^{0}}$.
Here, the third line follows from the inequalities $\delta_i \leq \frac{\ln(3/2)}{N}$ and $\left| \left(\bigotimes_{i=1}^N \bra{v_i^{j_i}} \right) \ket{\psi}\right| \leq \max_{\ket{\bm{u}} \in \mathrm{PS}_N} |\bra{\bm{u}} \psi \rangle|$, and the fifth line follows from the inequality $(1+x)^y \leq e^{xy}$ for positive $x,y$. 
From this inequality we can immediately obtain
\begin{equation}
    \max_{\ket{\bm{u}} \in \mathrm{PS}_N} |\bra{\bm{u}} \psi \rangle | \leq 2 \cdot |\langle \bm{v}^0|\psi \rangle | \leq 2 \cdot \max_{\ket{\bm{v}} \in \mathcal{B}_{N}} |\bra{\bm{v}} \psi \rangle |, 
\end{equation}
where the first inequality follows from Eq.~\eqref{eq:max_prod_ent_g}, and the second inequality follows from the fact that $\ket{\bm{v}^0}$ is an element of $\mathcal{B}_{N}$.
\end{proof}

\subsection{Tail bound for overlaps between a Haar-random state and a fixed state} \label{ss:tail_bound_Haar}

Here, we prove the following lemma, which provides an exponential tail bound for the squared overlap between a Haar-random pure state and a fixed pure state:
\begin{lemma}\label{lem:concent_Haar}
Let $\mu_{\rm H}$ be the uniform distribution over $D$-dimensional pure states, and let $\ket{\bm{v}}$ be a fixed pure state in the $D$-dimensional system.
Then, the following inequality is satisfied for the positive parameter $\alpha \geq 1$ and the constant $C_1 \equiv \frac{2}{9\pi^3 \ln 2}$:
\begin{equation}
\label{eq:lem2}
    \ProbH \left[\left| \langle \bm{v} \ket{\psi} \right|^2 \geq \frac{4\alpha}{D} \right] \leq 2  e^{-C_1 \alpha}.
\end{equation} 
\end{lemma}

The main tool for proving this lemma is the well-known Lévy's lemma, stated as follows:
\begin{lemma}[Lévy's lemma, Lem.~III.1 of Ref.~\cite{hayden2006aspects}] \label{lem:Levy's lemma}
Let $\mu_{\rm H}$ be the uniform distribution over $D$-dimensional pure states, and let $f$ be a function that maps a $D$-dimensional pure state to a real number, with Lipschitz constant $\eta$. Then, the following inequality is satisfied:
\begin{equation}
    \label{eq:Levy}
    \ProbH \left[ \left|f(\psi) - \Expsi[f(\psi)] \right| \geq \delta  \right] \leq 2 \exp\left(\frac{-2 D\delta^2}{9\pi^3 \ln 2 \eta^2} \right).
\end{equation}
\end{lemma}
\noindent
Using Lévy's lemma, we obtain Lemma~\ref{lem:concent_Haar} as follows:

\begin{proof}[Proof of Lemma~\ref{lem:concent_Haar}]
We prove Eq.~\eqref{eq:lem2} by applying Lévy's lemma to the function $f(\psi) \equiv \left| \langle \bm{v} \ket{\psi} \right|$. 
For this purpose, we evaluate its Lipschitz constant $\eta_f$ and the expectation value $\Expsi[f(\ket{\psi})]$.

First, we evaluate the Lipschitz constant.  
To this end, we choose an orthonormal basis $\{\ket{i}\}_{i=1}^D$ of the $D$-dimensional Hilbert space such that $\ket{1} = \ket{\bm{v}}$. 
Then, an arbitrary $D$-dimensional pure state can be written as
\begin{math}
    \ket{\psi} \equiv \sum_{i=1}^D \lambda_i^\psi \ket{i},
\end{math}
and the Lipschitz constant can be upper bounded as follows:
\begin{equation}
\label{eq:conc_2LOCC_prf1}
\begin{split}
    \eta_f &\equiv \sup_{\ket{\psi_1},\ket{\psi_2}} 
    \frac{|f(\ket{\psi_1}) - f(\ket{\psi_2})|}
    {\left\|\ket{\psi_1}-\ket{\psi_2} \right\|} \\
    &= \sup_{\{\lambda_i^{\psi_1}\},\{\lambda_i^{\psi_2}\}} 
    \frac{\left||\lambda_1^{\psi_1}|-|\lambda_1^{\psi_2}|\right|}
    {\sqrt{\sum_{i=1}^D |\lambda_i^{\psi_1}-\lambda_i^{\psi_2}|^2}} \\
    &\leq 1,
\end{split}
\end{equation}
where the second line follows from the identity $f(\ket{\psi}) = |\lambda^\psi_1|$, and the third line follows from the triangle inequalities  
$|\lambda_1^{\psi_1}| - |\lambda_1^{\psi_2}| \leq |\lambda_1^{\psi_1}-\lambda_1^{\psi_2}|$  
and  
$-|\lambda_1^{\psi_1}| + |\lambda_1^{\psi_2}| \leq |\lambda_1^{\psi_1}-\lambda_1^{\psi_2}|$.

Next, we evaluate the expectation value of $f(\ket{\psi})$:
\begin{equation}
\label{eq:conc_2LOCC_prf2}
\Expsi[ \left| \langle \bm{v} \ket{\psi} \right|]^2 
\leq 
\Expsi[  \left| \langle \bm{v} \ket{\psi} \right|^2] 
= \frac{1}{D},
\end{equation}
where the inequality follows from the convexity of the function $x^2$, and the equality is the well-known identity (see, e.g., Ref.~\cite{mele2024introduction}).

Using Eqs.~\eqref{eq:conc_2LOCC_prf1} and \eqref{eq:conc_2LOCC_prf2}, we obtain
\begin{align}
    \ProbH \left[ \left| \langle \bm{v} \ket{\psi} \right|^2 \geq \frac{4\alpha}{D} \right] 
        &= \ProbH \left[ \left| \langle \bm{v} \ket{\psi} \right| \geq 2\sqrt{\frac{\alpha}{D}} \right] \nonumber\\
        &\leq \ProbH \left[ 
        \left|\, 
        \left| \langle \bm{v} \ket{\psi} \right|
        - \Expsi[\left| \langle \bm{v} \ket{\psi} \right|]
        \right| \geq \sqrt{\frac{\alpha}{D}}
        \right] \nonumber \\
        &\leq 2 e^{-C_1 \alpha},
\end{align}
where the second line follows from Eq.~\eqref{eq:conc_2LOCC_prf2} and the assumption that $\alpha\geq 1$, and the third line follows from Lévy's lemma together with the bound for the Lipschitz constant given in Eq.~\eqref{eq:conc_2LOCC_prf1}. 
\end{proof}

\subsection{Proof of Theorem~\ref{thm:Haar_2LOCC}} \label{ss:prf_thm1}

Here, we provide the proof of Theorem~\ref{thm:Haar_2LOCC} by utilizing Lemmas~\ref{lem:e_net_g_ent} and~\ref{lem:concent_Haar}, together with Eq.~\eqref{eq:W_LOCC_UB}:

\begin{proof}[Proof of Theorem~\ref{thm:Haar_2LOCC}]
From Lemmas~\ref{lem:e_net_g_ent} and~\ref{lem:concent_Haar}, we can derive the following inequality:
\begin{align}
\ProbH \left[ W_{\rm LOCC} (\ket{\psi}) \geq  2 \ln N \right]
&\leq \ProbH \left[ E_g(\ket{\psi}) \leq N\ln d - 2 \ln N \right] \nonumber \\
&=\ProbH \left[ \max_{\ket{\bm{u}}\in \mathrm{PS}_N }  |\langle \bm{u} \ket{\psi} |^2  \geq  \frac{N^2}{d^N} \right] \nonumber \\
&\leq \ProbH \left[ \max_{\ket{\bm{v}}\in \mathcal{B}_N } |\langle \bm{v} \ket{\psi} |^2  \geq \frac{N^2}{4d^N} \right]  \nonumber\\
&\leq \sum_{\ket{\bm{v}}\in \mathcal{B}_N} \ProbH \left[\left|\langle \bm{v} \ket{\psi} \right|^2  \geq \frac{N^2}{4d^N} \right]  \nonumber\\
&\leq e^{ 2d N (\ln N +3)} \cdot 2 e^{-\frac{C_1N^2}{16}} \nonumber \\
&= e^{-O(N^2)}  \label{eq:thm1_proof}
\end{align}
where the first line follows from Eq.~\eqref{eq:W_LOCC_UB}, 
the second line is obtained by exponentiating both sides, 
the third line follows from Lemma~\ref{lem:e_net_g_ent}, 
the fourth line follows from the union bound, 
and the fifth line follows from Lemmas~\ref{lem:e_net_g_ent} and~\ref{lem:concent_Haar}. 
\end{proof}

\section{Extractable work from approximate state designs under LOCC} \label{s:t_design}

In this section, we prove Theorem~2 in the main text, which analyzes the extractable work for states sampled from approximate state designs. 
First, we introduce the definition of approximate state designs: 
\vspace{-2mm}
\begin{definition}[$\varepsilon$-approximate state $t$-design] \label{def:t_design}
An ensemble of pure states $\nu$ is an $\varepsilon$-approximate state $t$-design if the following operator inequality is satisfied:
\begin{equation}
\label{eq:approx_t_design}
    (1-\varepsilon) \Phi_{\mu_{\rm H}}^{(t)}\leq  \Phi_{\nu}^{(t)} \leq (1+\varepsilon) \Phi_{\mu_{\rm H}}^{(t)}.
\end{equation}
where $\Phi_{\nu}^{(t)} \equiv \Exnu \left[\ket{\psi}\bra{\psi}^{\otimes t} \right]$ is the $t$-th moment operator of an ensemble $\nu$, $\mu_{\rm H}$ is the uniform (Haar) distribution over the pure states, and $A\leq B$ for Hermitian operators means that $(B-A)$ is positive semidefinite.
\end{definition}
\vspace{-1.5mm}
\noindent
While there are several definitions of approximate state designs depending on how to quantify the difference between $\nu$ and Haar measure, we use Definition~\ref{def:t_design} throughout this manuscript.

Based on this definition, we derive the following theorem:
\vspace{-2mm}
\begin{theorem}[Theorem 2 in the main text]\label{thm:t-design}
Let $\nu$ be an $\varepsilon$-approximate state $t$-design on $N$-qudit pure states with $\varepsilon=O(1)$, and let $d$ be the dimension of each qudit. Then, the following statements hold:
\begin{enumerate}
    \item[$\mathrm{(i)}$] if $t\geq N$, we have the inequality
    \begin{equation}
    \label{eq:work_O(N)_design_adaptive}
        \Probnu \left[ W_{\rm LOCC} (\ket{\psi}) \geq (2d+5) \ln N  \right] \leq e^{-O(N\ln N)}.
    \end{equation}
    \item[$\mathrm{(ii)}$] if $t< N $, we have the inequality
    \begin{equation}
    \label{eq:work_o(N)_design_adaptive}
        \Probnu \left[ W_{\rm LOCC} (\ket{\psi}) \geq (2d+5) t^{-1} N \ln N  \right] \leq e^{-O(N\ln N)}.
    \end{equation}
\end{enumerate}
\end{theorem}
\vspace{-1.5mm}
\noindent
The second case $\mathrm{(ii)}$ of this theorem immediately implies that for $t = \omega (\ln N)$, the extractable work is not extensive, i.e., $W_{\rm LOCC}=o(N)$ with high probability.
The proof sketch of this theorem is similar to that of Theorem~\ref{thm:Haar_2LOCC}. 
To be specific, only the difference is the derivation of the tail bound. In the proof of Theorem~\ref{thm:t-design}, we use Lemma~\ref{lem:concent_tdesign}, which is introduced later in Sec.~\ref{ss:tail_t_design}, whereas in the proof of Theorem~\ref{thm:Haar_2LOCC}, we used Lemma~\ref{lem:concent_Haar}.

\subsection{Tail bound derived with $t$-design property} \label{ss:tail_t_design}

In this subsection, we introduce the following lemma, which provides an exponential tail bound for the squared overlap between a state sampled from $t$-design and a fixed product state.
\begin{lemma}\label{lem:concent_tdesign}
Let $\nu$ be an ensemble of $N$-qudit pure states which is an $\varepsilon$-approximate state $t$-design, and let $\ket{\bm{v} }$ be a fixed product state of $N$-qudit system.
Then, we have the following inequality for arbitrary $\alpha >0$:
\begin{equation}
\label{eq:lem4}
    \Probnu \left[ \left| \langle \bm{v} \ket{\psi} \right|^2 \geq \frac{\alpha}{D} \right] \leq  \left(\frac{t}{\alpha}\right)^t \cdot (1+\varepsilon),
\end{equation} 
where $D\equiv d^N$ is the dimension of $N$-qudit system. 
\end{lemma}

\begin{proof}
From the $t$-design property, we can obtain the following inequality: 
\begin{align}
\Exnu \left[ \left| \langle \bm{v} \ket{\psi} \right|^{2t} \right] &= {}^{\otimes t}\!\bra{\bm{v}} \Phi_{\nu}^{(t)}\ket{\bm{v}}^{\otimes t}\nonumber\\
&\leq {}^{\otimes t}\!\bra{\bm{v}} (1+\varepsilon) \Phi_{\mu_{\rm H}}^{(t)}\ket{\bm{v}}^{\otimes t} \nonumber\\
&= (1+\varepsilon) \cdot \frac{(D-1)!}{(t+D-1)!}  \sum_{\pi \in S_{t}} {}^{\otimes t}\bra{\bm{v}}V_{D} (\pi) \ket{\bm{v}}^{\otimes t}\nonumber\\
&= (1+\varepsilon) \cdot \frac{(D-1)!\ t!}{(t+D-1)!} \nonumber\\
&\leq (1+\varepsilon) \cdot \left(\frac{t}{D}\right)^t, \label{eq:t_design_concentration}
\end{align}
where $S_{t}$ denotes the set of permutations of $t$ elements, and $V_D(\pi)$ denotes the unitary representation of the permutation $\pi \in S_t$ acting on the $t$-fold tensor product Hilbert space $(\mathbb{C}^D)^{\otimes t}$, defined by
\[
V_D(\pi)\ket{i_1,i_2,\dots,i_t} = \ket{i_{\pi^{-1}(1)}, i_{\pi^{-1}(2)}, \dots, i_{\pi^{-1}(t)}}.
\]
The second line follows from Eq.~\eqref{eq:approx_t_design}, 
the third line follows from the well-known expression for the $t$-th moment operator of the Haar distribution~\cite{mele2024introduction},
\[
\Phi_{\mu_{\rm H}}^{(t)} = \sum_{\pi \in S_{t}} \frac{(D-1)!}{(t+D-1)!}\, V_{D} (\pi),
\]
and the fourth line follows from the fact that ${}^{\otimes t}\!\bra{\bm{v}} V_{D}(\pi) \ket{\bm{v}}^{\otimes t} = 1$ for any $\pi \in S_t$.

By further utilizing Markov's inequality $M \mathrm{Prob}_x [f(x) > M] \leq \mathbb{E}_x[f(x)]$ for positive function $f$, we can derive the following inequality:
\begin{align}
    \Probnu \left[ \left| \langle \bm{v} \ket{\psi} \right|^2 \geq \frac{\alpha}{D}\right] &= \Probnu \left[ \left| \langle \bm{v} \ket{\psi} \right|^{2t} \geq \left(\frac{\alpha}{ D}\right)^t \right]  \nonumber\\
    &\leq  \left(\frac{D}{\alpha}\right)^t  \Exnu \left[ \left| \langle \bm{v} \ket{\psi} \right|^{2t} \right] \nonumber\\
    &\leq  \left(\frac{t}{\alpha}\right)^t \cdot (1+\varepsilon),
\end{align}
where the final line follows from Eq.~\eqref{eq:t_design_concentration}.
\end{proof}

\subsection{Proof of Theorem~\ref{thm:t-design}}

By utilizing Lemmas~\ref{lem:e_net_g_ent} and \ref{lem:concent_tdesign}, we can prove Theorem~\ref{thm:t-design}.

\begin{proof}[Proof of Theorem~\ref{thm:t-design}]
When $\nu$ is an $\varepsilon$-approximate state $t$-design, we get the following inequality:
\begin{align}
\Probnu \left[ W_{\rm LOCC} (\ket{\psi}) \geq   \ln \alpha \right]
&\leq \Probnu \left[ E_g(\ket{\psi}) \leq N\ln d - \ln \alpha \right] \nonumber \\
&= \Probnu \left[ \max_{\ket{\bm{u}}\in \mathrm{PS}_N } \left|\langle \bm{u} \ket{\psi} \right|^2 \geq \frac{\alpha}{d^N} \right] \nonumber\\
&\leq \Probnu \left[ \max_{\ket{\bm{v}}\in \mathcal{B}_{N}} \left|\langle \bm{v} \ket{\psi} \right|^2 \geq \frac{\alpha}{4d^N} \right] \nonumber\\
&\leq \sum_{\ket{\bm{v}}\in \mathcal{B}_{N}} \Probnu \left[ \left|\langle \bm{v} \ket{\psi} \right|^2 \geq \frac{\alpha}{4d^N} \right] \nonumber\\
&\leq e^{2dN(\ln N+3)} \cdot \left(\frac{4t}{\alpha}\right)^t \cdot (1+\varepsilon) \nonumber\\
&\leq (1+\varepsilon)  \cdot \exp \left[ (3+\ln N) \cdot 2dN + 4 t \ln t - t \ln \alpha \right], \label{eq:prob_t_design_bound}
\end{align}
where the third line follows from Lemma~\ref{lem:e_net_g_ent}, the fourth line is derived by using the union bound, and the fifth line follows from Lemmas~\ref{lem:UB_cardinality_e-net} and \ref{lem:concent_tdesign}.

In the case of $\mathrm{(i)}$, we obtain the following inequality by substituting $t=N$ and $\alpha = N^{2d +5}$ in Eq.~\eqref{eq:prob_t_design_bound}:
\begin{align}
    \Probnu \left[ W_{\rm LOCC} (\ket{\psi}) \geq   (2d +5) \ln N \right]  &\leq  (1+\varepsilon)  \cdot \exp \left[ (3+\ln N) \cdot 2dN + 4 N \ln N - (2d +5) N \ln N \right]\nonumber\\
    &=  (1+\varepsilon)  \cdot e^{-N \ln N + 6dN} = e^{-O(N\ln N)}.
\end{align}
On the other hand, in the case of $\mathrm{(ii)}$, we obtain the following inequality by substituting $\alpha = N^{\frac{(2d+5) N}{t}}$ in Eq.~\eqref{eq:prob_t_design_bound}:
\begin{align}
    \Probnu \left[ W_{\rm LOCC} (\ket{\psi}) \geq   (2d +5) t^{-1} N\ln N \right]  &\leq  (1+\varepsilon)  \cdot \exp \left[ (3+\ln N) \cdot 2dN + 4 t \ln t - (2d +5)  N \ln N \right]\nonumber\\
    &\leq  (1+\varepsilon)  \cdot e^{-N \ln N + 6dN} = e^{-O(N\ln N)},
\end{align}
where we used $t<N$ in the derivation of the second line. 

\end{proof}

\section{Extractable work from graph states} \label{s:graph}

\subsection{Extractable work from random graph states} \label{ss:random_graph}
In this subsection, we provide the proof of Theorem~3 in the main text, which evaluates the scaling of extractable work from random graph states.
\vspace{-2mm}
\begin{theorem}[Theorem~3 in the main text] \label{thm:random_graph}
For the uniform distribution over $N$-vertex graphs $\mu_{\rm G}$, we have
\begin{equation}
\label{eq:W_LOCC_random_graph}
    \ProbmuG \!\left[W_{\rm LOCC}(\ket{G}) \geq c \sqrt{N}\ln N \right] \leq e^{-O(\sqrt{N})},
\end{equation}
where $c$ is a constant independent of $N$.
\end{theorem}
\vspace{-1.5mm}
\noindent
This theorem follows directly from the lemma below, proved in Ref.~\cite{GhoshPRXQ2025}.
\vspace{-2mm}
\begin{lemma}[Theorem~4 of Ref.~\cite{GhoshPRXQ2025}] \label{lem:Ghosh_PRXQ2025}
For the uniform distribution over $N$-vertex graphs $\mu_{\rm G}$, the following bound holds:
\begin{equation}
    \ProbmuG \!\left[E_g(\ket{G}) \leq N \ln 2 - c \sqrt{N}\ln N \right] \leq e^{-O(\sqrt{N})},
\end{equation}
where $E_g$ denotes the geometric entanglement, and $c$ is a constant independent of $N$.
\end{lemma} 
\vspace{-1.5mm}
\noindent

\begin{proof}[Proof of Theorem~\ref{thm:random_graph}]
Equation~\eqref{eq:W_LOCC_random_graph} follows from
\begin{align}
    \ProbmuG \!\left[ W_{\rm LOCC}(\ket{G}) \geq c \sqrt{N}\ln N \right]
    &\leq \ProbmuG \!\left[E_g(\ket{G}) \leq N \ln 2 - c \sqrt{N}\ln N \right]  \nonumber\\
    &\leq e^{-O(\sqrt{N})},
\end{align}
where the first inequality follows from Eq.~\eqref{eq:W_LOCC_UB}, and the second inequality follows from Lemma~\ref{lem:Ghosh_PRXQ2025}.
\end{proof}

\subsection{Extractable work from constant-degree graph states}

In this subsection, we prove the following proposition, which establishes the extensivity of work extraction from the graph states with a constant degree $r$:
\vspace{-2mm}
\begin{proposition}[Eq.~(11) in the main text] \label{prop:d_reg_graph}
For any graph states on $N$-qubit systems with a constant degree $r$, denoted by $\ket{G}$, the amount of extractable work under LOCC is lower bounded as
\begin{equation}
\label{eq:w_ext_d_reg}
    W_{\rm LOCC} (\ket{G}) \geq \frac{N}{r+1} \ln 2.
\end{equation}
\end{proposition}
\vspace{-1.5mm}
\noindent
We prove this proposition by explicitly constructing a work extraction protocol that achieves the above lower bound.
As preparation for the proof, we introduce the notions of the adjacency matrix and the independent set of a graph.

The adjacency matrix of an $N$-vertex graph $G = (V,E)$ is the $N \times N$ upper triangular matrix defined as
\begin{equation}
A_{ij} =
\begin{cases}
    1 & \mathrm{if\ } (i,j) \in E \ \mathrm{and}\ i<j, \\
    0 & \mathrm{otherwise},
\end{cases}
\end{equation}
which characterizes the connectivity of the graph.  
Using this matrix, the corresponding graph state can be written as
\begin{align}
    \ket{G} 
    &\equiv \left( \prod_{(i,j)\in E} CZ_{ij} \right)  \bigotimes_{n\in V} \ket{+}_n \nonumber \\
    &= \frac{1}{\sqrt{2^N}} \sum_{\substack{z_n\in \{0,1\} \\ \mathrm{for\ } n \in V}}  
    (-1)^{\sum_{i=1}^N \sum_{j=1}^{N} A_{ij} z_i z_j} 
    \bigotimes_{n\in V} \ket{z_n}_n,
    \label{eq:graph_rep_adj_mat}
\end{align}
where $\ket{z_n}_n$ for $z_n\in \{0,1\}$ denotes the computational-basis state of qubit $n$.

Next, we introduce the definition of the independent set.
\vspace{-2mm}
\begin{definition}[Independent set of graph]
For a graph $G=(V,E)$, a subset $S \subset V$ is called an independent set if no two vertices in $S$ are connected by an edge, i.e., $\forall i,j \in S$, $(i,j) \notin E$.
\end{definition}
\vspace{-1.5mm}
\noindent
With this definition, the exponent appearing in Eq.~\eqref{eq:graph_rep_adj_mat} can be rewritten as
\begin{equation}
\label{eq:adj_mat_ind_set}
    \sum_{i,j} A_{ij} z_i z_j 
    = \sum_{(i,j) \in E} z_i z_j
    = \sum_{i \in S} \left(\sum_{(i,j) \in E} z_j\right) z_i 
      + \sum_{\substack{(i,j) \in E \\ i \notin S,\ j \notin S}} z_i z_j,
\end{equation}
where the first equality follows from the definition of the adjacency matrix $A$, and the second follows from the definition of the independent set $S$.

Furthermore, regarding the size of independent sets, the following lemma is well known~\cite{caro1979new,wei1981lower}:
\vspace{-2mm}
\begin{lemma} \label{lem:caro_wei_bound}
For any $N$-vertex graph with fixed degree $r$, denoted by $G=(V,E)$, there exists an independent set $S$ satisfying
\begin{equation}
\label{eq:caro_wei_bound}
    |S| \geq \frac{N}{r+1}.
\end{equation}
\end{lemma}

We now prove Proposition~\ref{prop:d_reg_graph}.
\vspace{-2mm}
\begin{proof}[Proof of Proposition~\ref{prop:d_reg_graph}]
We derive Eq.~\eqref{eq:w_ext_d_reg} by constructing a one-round measurement protocol $\Lambda$, defined as follows:
\begin{itemize}
    \item Choose an independent set $S$ of the graph $G$.  
    By Lemma~\ref{lem:caro_wei_bound}, we can take $S$ so that $|S| \geq \frac{N}{r+1}$ holds.
    \item Perform projective measurements in $\{\ket{0}_n,\ket{1}_n\}$ on each qubit $n \in V \setminus S$, and in $\{\ket{+}_n,\ket{-}_n\}$ on each qubit $n \in S$, where $\ket{\pm}_n \equiv (\ket{0}_n \pm \ket{1}_n)/\sqrt{2}$.
\end{itemize}
\noindent
In this proof, we denote the measurement outcome of qubit $n$ by $y_n$, 
and the full outcome string by $y \equiv \{y_n\}_{n=1}^N$, 
omitting the superscript $1$ that is used elsewhere in this manuscript.
For $n \in V \setminus S$, the possible outcomes are $y_n = 0,1$, and for $n \in S$, they are $y_n = +,-$.

Then, the probability of obtaining outcome $y$ under protocol $\Lambda$ is given by 
\begin{align}
    P_{\Lambda,\ket{G}}[y]
    &= |\bra{y}G\rangle|^2 \nonumber\\
    &= \left| 
    \left(\bigotimes_{n\in V} {}_n\!\bra{y_n}\right)
    \left(\frac{1}{\sqrt{2^N}} 
    \sum_{\substack{z_n\in \{0,1\} \\ n \in V}}  
    (-1)^{\sum_{i,j} A_{ij} z_i z_j}
    \bigotimes_{n\in V} \ket{z_n}_n
    \right)
    \right|^2 \nonumber\\
    &= \frac{1}{2^{N}}
    \left| 
    \left(\bigotimes_{n\in S} {}_n\!\bra{y_n}\right)
    \left(
    \sum_{\substack{z_n\in \{0,1\}\\ n\in S}}
    (-1)^{\sum_{i\in S} (\sum_{(i,j)\in E} y_j) z_i}
    \bigotimes_{n\in S} \ket{z_n}_n
    \right)
    \right|^2 \nonumber\\
    &= \frac{1}{2^{N}}
    \prod_{n\in S}
    \left|
    \sum_{z_n\in \{0,1\}} 
    (-1)^{(\sum_{(n,j)\in E} y_j) z_n}
    {}_n\!\bra{y_n} z_n\rangle_n
    \right|^2 \nonumber\\
    &= \frac{1}{2^{N}} 
    \prod_{n\in S}
    \delta\!\left(y_n ; (-1)^{\sum_{(n,j)\in E} y_j}\right),
    \label{eq:reg_graph_prob}
\end{align}
where $\delta(a;b)$ is defined for $a\in\{+,-\}$ and $b\in\{\pm1\}$ as
\begin{equation} \label{eq:delta_def}
\delta(a;b)=
\begin{cases}
    2 & \mathrm{if\ } (a,b) = (+,1)\ \mathrm{or}\ (a,b)=(-,-1), \\
    0 & \mathrm{otherwise}.
\end{cases}
\end{equation}

From Eqs.~\eqref{eq:reg_graph_prob} and \eqref{eq:delta_def}, the probability distribution is uniform over exactly $2^{(N-|S|)}$ valid outcome strings, each with probability $2^{-(N-|S|)}$, and zero otherwise.  
Therefore, the Shannon entropy of the distribution is $H_{\Lambda,\ket{G}}(y) = (N-|S|)\ln 2$, and Eq.~\eqref{eq:w_ext_d_reg} is obtained as 
\begin{align*}
    W_{\rm LOCC}(\ket{G})
    &\geq N\ln 2 - H_{\Lambda,\ket{G}}(y) \\
    &\geq \frac{N}{r+1}\ln 2,
\end{align*}
where the final inequality uses $|S| \geq \frac{N}{r+1}$.
\end{proof}

\bibliography{bibliography}